# A Tutorial of 802.11 Implementation in NS-2

Yue Wang[1], On-Ching Yue[2]
1. School of Information, Central University of Finance and Economics
   Email: yue.l.wang@gmail.com
2. Department of Information Engineering, The Chinese University of Hong Kong
   Email: onching@ie.cuhk.edu.hk

 (Version 2, revised at July 20, 2007)

**Abstract:** By analyzing the source codes of ns-2, we discuss the simulated implementations of wireless channels, network interfaces and mostly the 802.11 MAC protocol in ns-2. We also notice the "bugs" of the 802.11 simulation compared with the reality, and present an extension to fading channels as well.

## 1. Introduction to ns-2

### 1.1 ns-2

Ns-2 [1] is a packet-level simulator which is essentially a *centralized* discrete event scheduler to schedule the events such as packet and timer expiration. The centralized event scheduler cannot accurately emulate "events occurred at the same time", instead, it can only handle events occurred one by one in time. However, this is not a serious problem in most network simulations, because the events here are often transitory. Besides, ns-2 implements a variety of network components and protocols. Notably, the wireless extension, derived from CMU Monarch Project [2], has 2 assumptions simplifying the physical world:
    (1) Nodes do not move significantly over the length of time they transmit or receive a packet. This assumption holds only for mobile nodes of high-rate and low-speed. Consider a node with the sending rate of 10Kbps and moving speed of 10m/s, during its receiving a packet of 1500B, the node moves 12m. Thus, the surrounding can change significantly and cause reception failure.
    (2) Node velocity is insignificant compared to the speed of light. In particular, none of the provided propagation models include Doppler effects, although they could.

### 1.2 GloMoSim and OPNET

GloMoSim [3] is another open-source network simulator which is based on parallel programming.

Hopefully, it can emulate the real world more accurately. However, it may be hard to debug parallel programs. Although GloMoSim currently solely supports wireless networks, it provides more physical-layer models than ns-2. There is another simulator OPNET which requires licence. Table 1 [4] compares the wireless physical models used in the three simulators.

Table 1. Physical layer and propagation models available in GloMoSim, ns-2 and OPNET

| Simulator | GloMoSim | ns-2 | OPNET |
|---|---|---|---|
| Noise (SNR) calculation | Cumulative | Comparison of two signals | Cumulative |
| Signal reception | SNRT based, BER based | SNRT based | BER based |
| Fading | Rayleigh, Ricean | Not included* | Not included |
| Path loss | Free space, Two ray, etc. | Free space, Two ray | Free space |

## 1.3 Ns-2 Basics

### Ns-2 directory structure

As shown in Figure 1, the C++ classes of ns-2 network components and protocols are implemented in the subdirectory "ns-2.*", and the TCL library (corresponding to configurations of these C++ instances) in the subdirectory of "tcl".

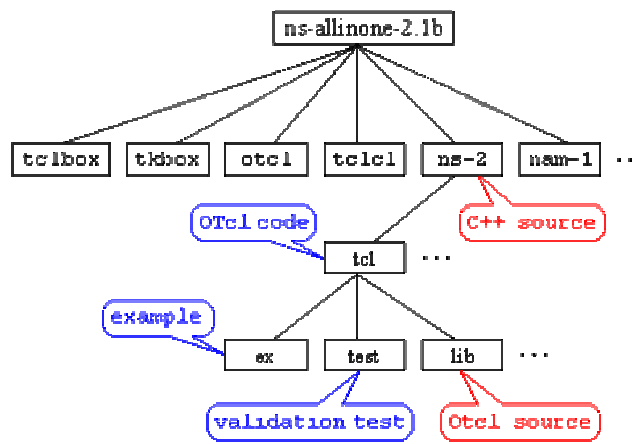

Figure 1. Ns-2 directory structure

## Network Components

Network components are Node, Link, Queue, etc. Some of them are simple components, that is, they corresponds to a single C++ object; The others are compound components, that is, they combine multiple simple components, e.g. a Link component is composed of a Delay component (emulating propagation delay) and a Queue component. In general, all network components are created, plugged and configured by some TCL scripts (./ns-allinone-2.*/tcl/) when ns-2 is initialized.

Example: Plug MAC into NetIF (Network Interface)
class MAC {
    void send (Packet* p);
    void recv(Packet*, Handler* h);
    NsObject* target_ //pointing to an instance of NetIF
}

## Event Scheduling

Events are something associated with time. class Event is defined by {time, uid, next, handler}, where time is the scheduling time of the event, uid is the event's id , next points to the next scheduling event in the event queue, and handler points to the function to handle the event at scheduling time. Events are put into the event queue sorted by their time, and scheduled one by one by the event scheduler. Note that class Packet is a subclass of class Event to simulate packets transmitted or received at some time. All network components are subclasses of class Handler as they need to handle events such as packets.

The scheduling procedure (void Scheduler::schedule(Handler* h, Event* e, double delay)) is shown in Figure 2. The event at the head of the event queue is delivered to its hander of some network component (object). Then, this network object may call other network objects to further handle this event, and may generate new events which are inserted into the event queue.

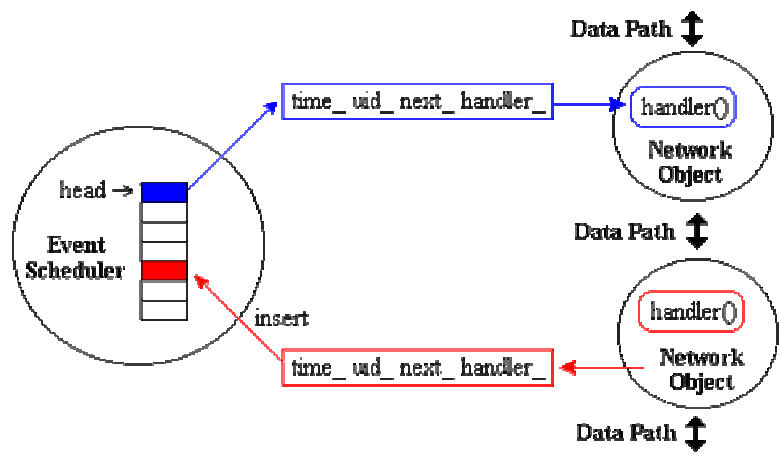

Figure 2. Discrete Event Scheduler

Example 1: A and B are two mobile nodes within the tx range. A sends packet *p* to B.

    A::send (Packet* p) {target_->recv(p)} //target_ points to B and will call B::recv

    B::recv(Packet*, Handler* h = 0) {

        …

        //target_ is B; schedule p at the time of (current_time + tx_time)

        Scheduler::instance().schedule(target_, p, tx_time)

        …

}

Example 2: Timer is another kind of Event which is handled by TimerHandler

class TimerHandler: public Handler

    void resched(double delay) //set the timer's expiration time, i.e. current_time + delay

    void handle(Event *e){//handler function at expiration time

        expire (Event *e) //the virtual handling function that needs to be overloaded by users

    }

**Note:**    There are *no real* time, timer, and packet tx/rx in ns-2 as in UNIX network programming.

## 2. 802.11 Implementation

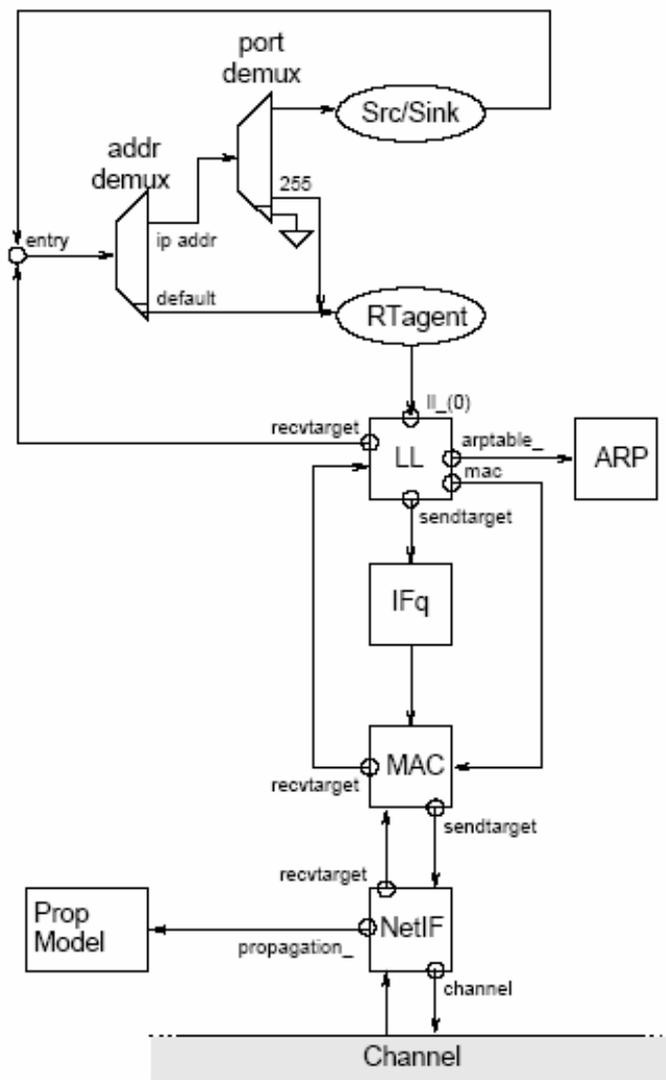

**Figure 3 Schematic of a mobile node under the CMU Monarch wireless extensions to ns.**

This section is a tutorial of source code analysis for ns-2.30. Please refer the Appendixes for more information. Figure 3 shows the network components in the mobile node and the data path of sending and receiving packets.

## 2.1 Physical Layer

### Channel

The function of class WirelessChannel (see *channel.cc*) is to deliver packets from a wireless node to its neighbors within *sensing range*, i.e.

    distCST_ = wifp->getDist(wifp->getCSThresh(), wifp->getPt(), 1.0, 1.0,
                highestZ , highestZ, wifp->getL(),   wifp->getLambda());

Note: distCST is calculated by the parameters such as CS Threshold, transmission power, antenna gains, antenna heights, system loss factor, and wavelength of light.

## NetIF (WirelessPhy)

The function of class WirelessPhy is to send packets to Channel and receive packet from Channel (see *wireless-phy.cc*).

(1) Packet Sending, sendDown()
   First, stamp txinfo in the packet before sending, i.e.
   p->txinfo_.stamp((MobileNode*)node(), ant_->copy(), Pt_, lambda_).
Here node() are the pointer of the sending node, ant_->copy() is the antenna's parameters such as the height of the antenna, Pt_ is the transmitting power, and lamba_ is the wavelength of light. These information is used for the receiving node to calculate the receiving power.
   Second, delivery the packet to channel, i.e. channel_->recv(p, this);

(2) Packet Reception, sendUp()
```
        //calculate Rx power by path loss models
        Pr = propagation_->Pr(&p->txinfo_, &s, this)

        if (Pr < CSThresh_) {
            pkt_recvd = 0; // cannot hear it
            …
        }
        if (Pr >= CSThresh_ && Pr < RXThresh_){
            pkt_recvd = 1;
            hdr->error = 1; // error reception, only for carrier sensing
            …
        }
        if (Pr >= RXThresh_) {
            pkt_recvd = 1;
            hdr->error = 0; // can be correct reception
            …
        }
```

First, ns-2 calculates the receiving power Pr by the tx_info_ of p and the receiver this. When Pr is less than CSThresh_ (carrier sense threshold), the receiver cannot sense the packet; Otherwise, the receiver can sense the packet and it can further decode (receive) it when Pr > RXThresh_ (i.e. reception threshold, which is > CSThresh_). Besides, the successful reception also depend on the packet's SNR is larger than CPThresh_ (capture threshold), note that which is checked in MAC layer.

## 2.2 MAC Layer (802.11)

class Mac802_11 has two functions: sending and receiving (see *mac-802_11.cc*). On sending, it uses CSMA/CA channel access mechanism; On receiving, it adopts a SNR threshold based capture model.

### State Transition Diagram & Main States

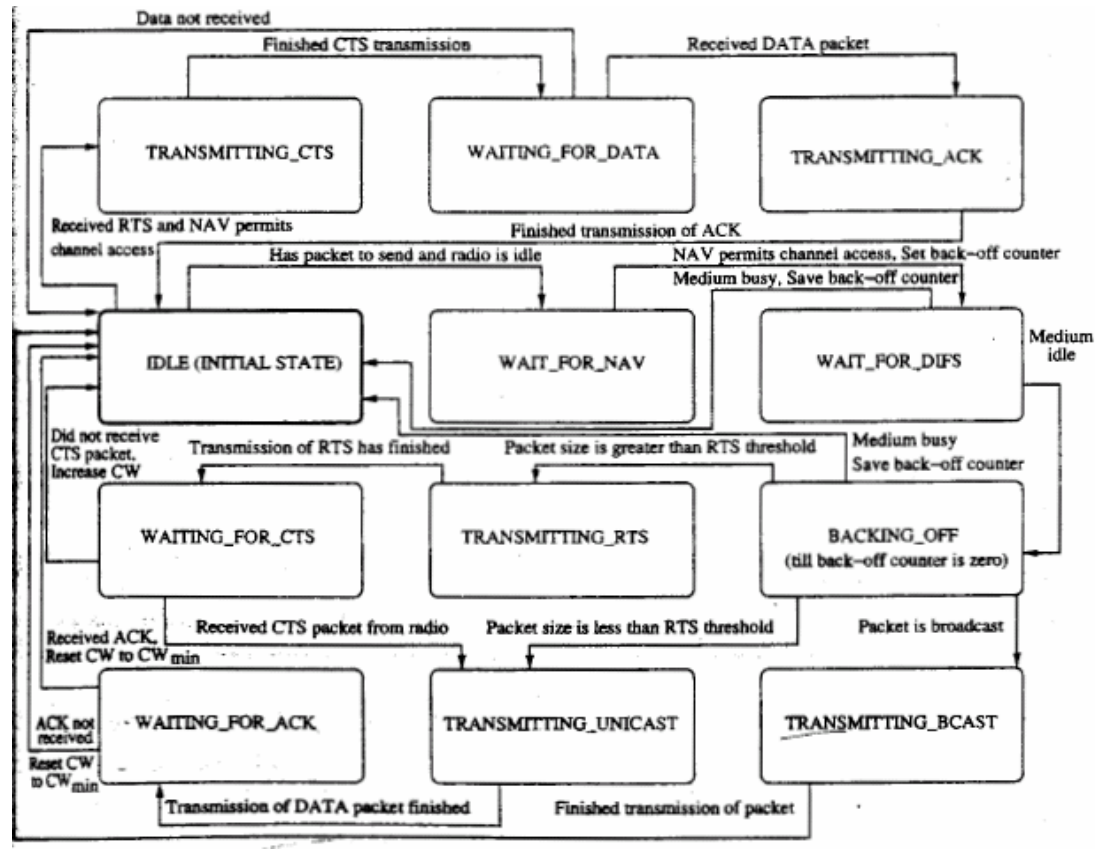

Figure 4.   802.11 MAC state transition diagram

State transition diagram can help us read or write network programs. Figure 4 shows a reference 802.11 state transition diagram [5]. The main states in ns-2 are described as follows.

```
enum MacState {
    MAC_IDLE    = 0x0000,
    MAC_POLLING = 0x0001, // ns 802.11 does not implement Polling
    MAC_RECV    = 0x0010,
    MAC_SEND    = 0x0100,
    MAC_RTS     = 0x0200,
    MAC_CTS     = 0x0400,
    MAC_ACK     = 0x0800,
    MAC_COLL    = 0x1000
};
```

```
MacState     rx_state_ //can be MAC_IDLE, MAC_RECV, MAC_COLL
MacState     tx_state_//can be MAC_IDLE, MAC_SEND, MAC_RTS, MAT_CTS, MAC_ACK
double       nav_ //expiration time of Network Allocation Vector

//channel is idle
int is_idle() {
   if (tx_state_ == MAC_IDLE && rx_state_ == MAC_IDLE && nav_ <= NOW)
        return 1;
   else
        return 0;
}
```

**Note:** is_idle() checks whether the channel is idle at the moment when it is called.

## MAC Timers

Timers are very important in 802.11 for triggering channel access. The following shows the basic timers and their functions.

BackoffTimer mhBackoff_
    void start(int cw, int idle);//when is_idle()==1, start to count down; otherwise freeze the timer
    void pause(); //freeze the timer when is_idle()==0
    void resume(double difs);//resume to count down when is_idle()==1 again
    void handle(Event *); //call backoffHandler to send RTS or DATA after it times out
    int busy(); //counting down

DeferTimer mhDefer_
    void start(double defer);//start to count down
    void handle (Event *);// send packets after it times out (eg. send CTS or ACK after SIFS expires)
    int busy(); //counting down

IFTimer      mhIF_;    // interface timer, interface state is active when transmitting
NavTimer     mhNav_;   // NAV timer
RxTimer      mhRecv_;  //completion of incoming packets; it will call recvHandler() {…
                       //recv_timer(); …}
TxTimer      mhSend_;  //sending timeout (e.g. no ACK received); it will call sendHandler() {
                       //… send_timer(); …}

**Note:** send_timer( ) resends RTS or DATA when mhSend_ times out.
    recv_timer() processes the received packet

The following invocation tree shows that show how receiving and sending will change MAC state and further control the backoff timer.

recv or send functions
    void setTxState (MacState newState) //For tx_state_
    void setRxState (MacState newState)//For rx_state_
        void checkBackoffTimer() {
            if(is_idle() && mhBackoff_.paused())
                mhBackoff_.resume(phymib_.getDIFS());
            if(!is_idle() && mhBackoff_.busy() && ! mhBackoff_.paused())
                mhBackoff_.pause();
        }

NAV timer is set by the received interfering packets to indicate the residual time of data transmission. The NAV timer is set for correctly decoded packets. Besides, in ns-2, capture( ) also updates the NAV timer so that this does not screw up carrier sense even when channel is idle. When NAV timer expires, navHandler() is called to resume backoff timer.

```
/* Note: nav_ expires also mean channel is idle, then call mhBackoff_resume() */
void set_nav(u_int16_t us) {
        double now = Scheduler::instance().clock();
        double t = us * 1e-6;
        if((now + t) > nav_) {
            nav_ = now + t;
            if(mhNav_.busy())
                mhNav_.stop();//reset nav_
            mhNav_.start(t);
        }
}
```

## CSMA/CA

recv function is generally the entry of most network protocols (invoked by both upper layer and lower layer). For outgoing packets, it will call send function which is the entry of CSMA/CA.

```
void recv(Packet *p, Handler *h) {
    struct hdr_cmn *hdr = HDR_CMN(p);
    //handle outgoing packets
    if(hdr->direction() == hdr_cmn::DOWN) {
                send(p, h); //CSMA/CA
                return;
        }
```

```
    …
    // handle incoming packets
    …
}

void send(Packet *p, Handler *h) {
  …
  if(mhBackoff_.busy() == 0) {
        if(is_idle()) {
            if (mhDefer_.busy() == 0) {
                /* If we are already deferring, there is no need to reset the Defer timer.*/
                mhBackoff_.start(cw_, is_idle(), phymib_.getDIFS());
            }
        } else {
            /* If the medium is NOT IDLE, then we start the backoff timer.*/
            mhBackoff_.start(cw_, is_idle());
        }
    }
}
```

**Note:** Packets will be transmitted in backoffHandler() after mhBackoff_ expires.

## Capture Model

Ns-2 uses a simplified capture model: When multiple packets collide at the receiver, only the *first* packet can be successfully received if its Rx Power should be larger than any of the other packets by at least CPThresh (10dB in ns-2).

```
void recv(Packet *p, Handler *h){
  …
//Handle incoming packets

/* When there is no packet reception, log receiving p at pktRx_*/
if(rx_state_ == MAC_IDLE) {
        setRxState(MAC_RECV);
        pktRx_ = p;
        mhRecv_.start(txtime(p)); // schedule the reception of this packet in txtime
}
/* When there is already a packet reception (in pktRx_), calculate the inference*/
else {
        //Simplified SNR calculation (only for two packets)
        if(pktRx_->txinfo_.RxPr / p->txinfo_.RxPr >= p->txinfo_.CPThresh) {
            capture(p);//pktRx_ can be correctly received;
```

```
        } else {
            collision(p);//stop receving pktRx_ (will call mhRecv.stop() )
        }
    }
}
```

## 2.3 Network Layer (AODV)

## 2.4 My Fading Extension

The probability density function of Rayleigh fading is $pdf(r) = \frac{r}{\sigma^2} e^{-\frac{r^2}{2\sigma^2}}$, where $r$ stands for Voltage. As power $P = c \cdot r^2$, its probability density function under Rayleigh fading is $pdf(P) = \frac{1}{\overline{P}} e^{-\frac{P}{\overline{P}}}$, where $\overline{P}$ is the mean of $P$ calculated by some path loss model. So, we can add fading extension after Pr ().

```
#include <random.h>
…
int WirelessPhy::sendUp(Packet *p) {

double Pr;
…
if (propagation_) {
                s.stamp((MobileNode*)node(), ant_, 0, lambda_);
                Pr = propagation_->Pr(&p->txinfo_, &s, this);

                /* Add Rayleigh fading (neglect time-correlation)*/
                double mean = Pr;
                Pr = Random::exponential(mean);

                if (Pr < CSThresh_)
                …
}
```

We do two experiments in 11Mbps 802.11 networks to see the impact of Rayleigh fading. The default tx power Pt is 0.28, and thus tx range and sensing range are 250m and 550m.

The first experiment is to test TCP performance (see Figure 5). We vary the distance $d$ from 50m to 250m. Figure 6 shows TCP throughputs as a function of $d$. When d becomes larger, fading can cause more packet loss and thus reduce TCP throughputs significantly.

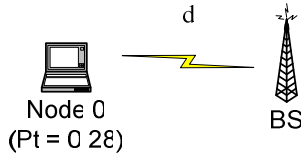

**Figure 5. TCP under Rayleigh fading (Node 0 sends TCP packets to BS )**

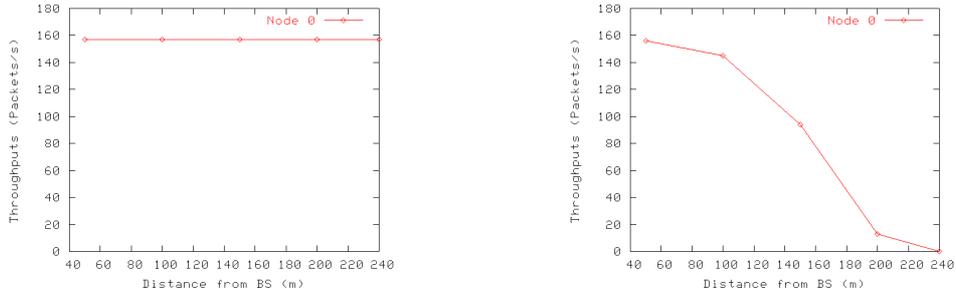

**Figure 6. TCP Throughputs a function of the distance from BS (a) Without Fading (b) Rayleigh Fading**

The second experiment (Figure 7) is to test UDP performance (assume saturate condition). We set Pt of Node 0 be 10 times of the default Pt and the SNR threshold is 10dB. Therefore, packets from Node 0 will be captured when they collide with Node 1 at BS suppose there is no fading. As shown in Figure 8, fading aggravates the unfairness of two senders when *d* becomes larger.

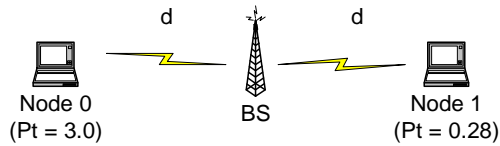

**Figure 7. Capture under Rayleigh fading (Node 0 and 1 send CBR packets to BS)**

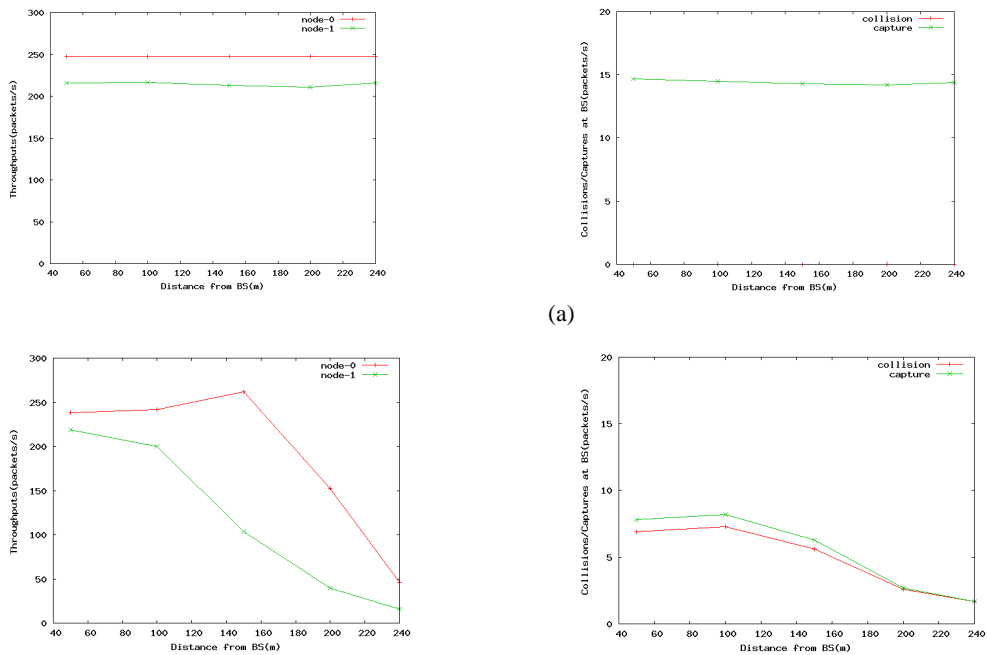

(a)

(b)

**Figure 8.  UDP Throughputs and Captures/Collisions at a function of the distance *d* from BS (a) Without Fading (b) Rayleigh Fading**

Finally, our simple implementation of fading does not consider time correlation. Please refer to CMU's patch for a more accurate fading extension [6].

## 2.5 "Bugs"

### Oversimplifications

Simulators always need simplifications to make their calculations viable (recall Section 1.1). However, when some assumptions are crucial in your simulations, you must be careful. For example, there is no scanning for WLAN (Discovery/Select/Authentication/Association) in ns-2, mobile nodes are associated with their BS automatically if they all have a same address prefix. If you want to study the overhead of scanning, you should make your extension.

### Standard Misinterpretation

Ns-2 may misinterpret some network protocol standards for it is an open project. For example, we find ns-2 802.11 implementation seems abuse EIFS (set_nav(usec(phymib_.getEIFS() + txtime(p))) // whenever p is error, defer EIFS) [7]. Actually, in Figure 7, the node with Pt 0.28 will has more throughput before we disable getEIFS().

## 3. Acknowledgement

I put this report in Arxiv for permanent storage, since many are interested and some even cited it in the past few years. I did this work when I was a research assistant and Ph.D student at Department of Computer Science & Engineering, The Chinese University of Hong Kong (CUHK) in 2005-2006. I appreciate Prof. On-Ching Yue very much for his nice guidance and suggestions, and appreciate Prof. John C. S. Lui and Prof. Dah-Ming Chiu for their supports as well ☺

# Appendix I: TCL lib for wireless nodes

In "tcl/lib/ns-mobile.tcl", **add-interface** set up physical layer, link layer, mac layer and network layer for wireless nodes. See:

*Node/MobileNode instproc add-interface { channel pmodel lltype mactype qtype qlen iftype anttype topo inerrproc outerrproc fecproc } {…}*

# Appendix II: 802.11 Key Functions

**tx_resume( ):** invoked after recv_RTS( ), recv_CTS( ), recv_DATA( ), recv_ACK( ), send_timer( ) for the subsequent transmission. If no packets holding in MAC layer, tx_resume( ) callback, that is, it fetches a packet from IFQ and then invokes recv( ) to send out this packet.

*h->handle( ); // h is a handler of IFQ*
*void QueueHandler::handle(Event\*) {*
 *queue_.resume( );*
*}*

**recvACK() -> tx_resume() -> rx_resume()**

**check_pktTx( ):** all DATA packets are transmitted in this function, which is invoked by the handler functions of mhDefer_ or mhBackoff, i.e. deferHandler() or backoffHandler().

**checkBackoffTimer( ):** resume or pause mhBackoff_, which is invoked by setRxState( ) or setTxState( )

**navHandler( ):** resume mhBackoff_

**setRxState( ):**
 **setRxState(MAC_RECV):** invoked by recv( )
 **setRxState(MAC_COLL):** invoked by collision( )
 **setRxState(MAC_IDLE):** invoked by rx_resume( ), which is invoked by recv_timer( )

**setTxState( ):**
    **setTxState( MAC_RTS):** invoked by check_pktRTS( )
    **setTxState( MAC_CTS), setTxState( MAC_ACK):** invoked by check_pktCTRL( )
    **setTxState( MAC_SEND):** invoked by check_pktTx( )
    **setTxState( MAC_IDLE):** invoked by tx_resume( )

# Appendix III:   Scenario Generation

In "indep-utils/cmu-scen-gen":
(1) Generate topology

*./setdes [parameters]*

(2) Generate TCP/CBR traffic

*ns cbrgen.tcl [parameters]*

Note: Seed cannot be 0; rate (in bps) must be a float value